\documentclass[sigconf,edbt]{acmart-edbt2025}

\def\BibTeX{{\rm B\kern-.05em{\sc i\kern-.025em b}\kern-.08em
    T\kern-.1667em\lower.7ex\hbox{E}\kern-.125emX}}

\usepackage{booktabs} 

\setcopyright{rightsretained}

\acmDOI{}

\acmISBN{XXXXXXXXXXXX}

\acmConference[EDBT 2025]{28th International Conference on Extending Database Technology (EDBT)}{25th March-28th March, 2025}{Barcelona, Spain}
\acmYear{2025}

\usepackage{balance}       

\usepackage{xcolor}
\usepackage{soul}
\usepackage{subcaption}

\usepackage{algorithm}
\usepackage{algorithmic}
\usepackage{pgfplots}
\usepackage{xcolor}
\definecolor{wdc_color}{RGB}{31, 119, 180}  
\definecolor{gds_color}{RGB}{255, 127, 14}  
\usetikzlibrary{matrix}
\usepgfplotslibrary{groupplots}
\pgfplotsset{compat=newest}
\usepackage{amsmath}
\usepackage{amsfonts}
\usepackage{graphicx}
\usepackage{color}
\usepackage{framed}

\pgfplotsset{compat=1.17}
\definecolor{lightgray}{gray}{0.9}

\AtBeginDocument{%
  \providecommand\BibTeX{{%
    Bib\TeX}}}

\begin{document}
\title{Gem: Gaussian Mixture Model Embeddings for Numerical Feature Distributions}

\author{Hafiz Tayyab Rauf} 
\orcid{0000-0002-1515-3187}
\affiliation{%
	\institution{Department of Computer Science, \\ University of Manchester}
	\city{Manchester}
	\country{UK.}
	\postcode{M13 9PL}
}
\email{hafiztayyab.rauf@manchester.ac.uk}

\author{Alex Bogatu} 
\orcid{}
\affiliation{%
	\institution{National Biomarker Centre, CRUK-MI, \\ University of Manchester}
	\city{Manchester}
	\country{UK.}
	\postcode{M13 9PL}
}
\email{alex.bogatu@manchester.ac.uk}

\author{Norman W. Paton}
\orcid{0000-0003-2008-6617}
\affiliation{%
	\institution{Department of Computer Science, \\ University of Manchester}
	\city{Manchester}
	\country{UK,}
	\postcode{M13 9PL}
}
\email{norman.paton@manchester.ac.uk}

\author{Andre Freitas}
\orcid{0000-0002-4430-4837}
\affiliation{%
	\institution{Department of Computer Science, \\ University of Manchester\\ Manchester, UK.\\ IDIAP Research Institute\\ Martigny, Switzerland.}
	\postcode{M13 9PL}
}
\email{andre.freitas@manchester.ac.uk}

\renewcommand{\shortauthors}{Rauf et al.}

\begin{abstract}
Embeddings are now used to underpin a wide variety of data management tasks, including entity resolution, dataset search and semantic type detection. Such applications often involve datasets with numerical columns, but there has been more emphasis placed on the semantics of categorical data in embeddings than on the distinctive features of numerical data. In this paper, we propose a method called Gem (\textit{G}aussian mixture model \textit{em}beddings) that creates embeddings that build on numerical value distributions from columns. The proposed method specializes a Gaussian Mixture Model (GMM) to identify and cluster columns with similar value distributions. We introduce a signature mechanism that generates a probability matrix for each column, indicating its likelihood of belonging to specific Gaussian components, which can be used for different applications, such as to determine semantic types. Finally, we generate embeddings for three numerical data properties: distributional, statistical, and contextual. Our core method focuses solely on numerical columns without using table names or neighboring columns for context. However, the method can be combined with other types of evidence, and we later integrate attribute names with the Gaussian embeddings to evaluate the method's contribution to improving overall performance. We compare Gem with several baseline methods for numeric only and numeric + context tasks, showing that Gem consistently outperforms the baselines on four benchmark datasets.
\end{abstract}



\maketitle

\section{Introduction}
\label{Intro}

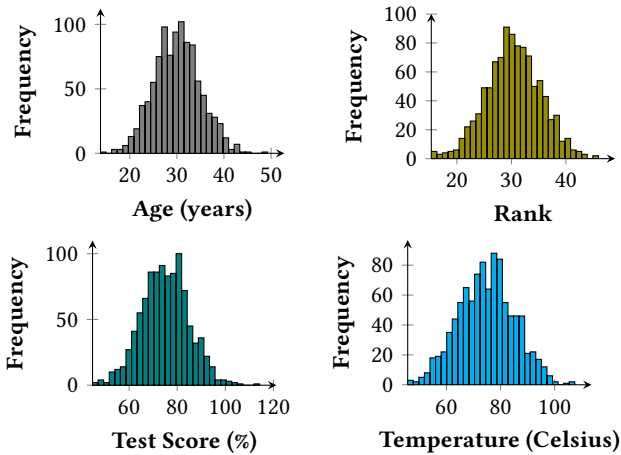
\begin{figure}[t]
    \centering
    \begin{tabular}{cc}
        \begin{tikzpicture}
        \begin{axis}[
            width=4cm,
            height=3.5cm,
            xlabel={\textbf{Age (years)}},
            ylabel={\textbf{Frequency}},
            yticklabel style={/pgf/number format/fixed},
            tick label style={font=\bfseries},
            label style={font=\bfseries},
            title style={font=\bfseries},
            every axis plot/.append style={unbold/.style={/pgfplots/mark options={solid}}},
            axis lines=left,
            enlargelimits=upper
        ]
        \addplot+[
            ybar,
            hist={
                bins=30,
            },
            fill=gray,
            draw=black,
            mark=none 
        ] table[y index=0] {age_data.dat};
        \end{axis}
        \end{tikzpicture} &
        
        \begin{tikzpicture}
        \begin{axis}[
            width=4cm,
            height=3.5cm,
            xlabel={\textbf{Rank}},
            ylabel={\textbf{Frequency}},
            yticklabel style={/pgf/number format/fixed},
            tick label style={font=\bfseries},
            label style={font=\bfseries},
            title style={font=\bfseries},
            every axis plot/.append style={unbold/.style={/pgfplots/mark options={solid}}},
            axis lines=left,
            enlargelimits=upper
        ]
        \addplot+[
            ybar,
            hist={
                bins=30,
            },
            fill=olive,
            draw=black,
            mark=none 
        ] table[y index=0] {rank_data.dat};
        \end{axis}
        \end{tikzpicture} \\
        
        \begin{tikzpicture}
        \begin{axis}[
            width=4cm,
            height=3.5cm,
            xlabel={\textbf{Test Score (\%)}},
            ylabel={\textbf{Frequency}},
            yticklabel style={/pgf/number format/fixed},
            tick label style={font=\bfseries},
            label style={font=\bfseries},
            title style={font=\bfseries},
            every axis plot/.append style={unbold/.style={/pgfplots/mark options={solid}}},
            axis lines=left,
            enlargelimits=upper
        ]
        \addplot+[
            ybar,
            hist={
                bins=30,
            },
            fill=teal,
            draw=black,
            mark=none 
        ] table[y index=0] {test_score_data.dat};
        \end{axis}
        \end{tikzpicture} &
        
        \begin{tikzpicture}
        \begin{axis}[
            width=4cm,
            height=3.5cm,
            xlabel={\textbf{Temperature (Celsius)}},
            ylabel={\textbf{Frequency}},
            yticklabel style={/pgf/number format/fixed},
            tick label style={font=\bfseries},
            label style={font=\bfseries},
            title style={font=\bfseries},
            every axis plot/.append style={unbold/.style={/pgfplots/mark options={solid}}},
            axis lines=left,
            enlargelimits=upper
        ]
        \addplot+[
            ybar,
            hist={
                bins=30,
            },
            fill=cyan,
            draw=black,
            mark=none 
        ] table[y index=0] {temperature_data.dat};
        \end{axis}
        \end{tikzpicture} \\
    \end{tabular}
    \caption{A histogram with a Kernel Density Estimate (KDE) overlay distributions of four numerical columns: Age, Rank, Test Score, and Temperature. Despite the similar distribution shapes — Age and Rank both showing a normal distribution around a mean of 30 and Test Score and Temperature around a mean of 75 — the semantic contexts differ significantly and refer to the different semantic types and units. For example, \textit{"Age"} might be measured in years, \textit{"Rank"} in a hierarchical position, \textit{"Test Score"} as points out of 100, and \textit{"Temperature"} in degrees, Fahrenheit and Celsius. These variations illustrate the complexity of semantic type detection of columns with different distributions. In this context, existing methods struggle to distinguish these overlapping columns. However, our proposal can effectively distinguish between these columns by focusing on their distributional properties.}
\label{fig:distribution}
\end{figure}

Data repositories, such as data lakes and open government data, often contain substantial amounts of numerical data \cite{DBLP:conf/edbt/LangeneckerSSB24}, which forms the backbone of various analytical and predictive models. Given its extensive use, numerical data often outnumbers non-numerical and categorical data \cite{DBLP:conf/edbt/LangeneckerSSB24}. Applications such as semantic type detection of numerical data are thus important, but numerical data presents several challenges, including variability in data distributions (e.g., consider two columns both labeled \textit{"weight"} in different datasets: one representing \textit{"package weight"}, and another representing \textit{"human weight"}. Although both columns share the same label and numerical nature, their distributions and contexts differ significantly). Even when columns have similar distributions, their semantics might differ. For instance, a column representing \textit{temperature} and another representing \textit{test score} can have similar distribution shapes but different semantics. Figure \ref{fig:distribution} illustrates the challenge faced when comparing numerical distributions, where the columns from different semantic types share similar values.

Although several approaches handle numerical columns using bespoke deep learning techniques \cite{DBLP:conf/emnlp/SundararamanSSW20,DBLP:conf/nips/GorishniyRB22,DBLP:conf/emnlp/JiangNGCZST20,DBLP:journals/pacmmod/LangeneckerSSB23,DBLP:conf/nips/KadraLHG21,DBLP:journals/pvldb/ZhangSLHDT20,DBLP:journals/pvldb/SunXC23}, they heavily rely on the context extracted from non-numerical data. For example, Pythagoras \cite{DBLP:conf/edbt/LangeneckerSSB24} creates a graph where nodes represent columns and edges denote relationships derived from the table's metadata and neighboring columns. Our contribution complements that of existing proposals, in providing a new approach to handling numerical features that can be combined with other features or used in isolation where contextual information is limited. Furthermore, existing methods often overlook the distributional differences between columns with similar column names but different values. For example, two columns representing temperature readings in different regions might have similar schemas but different distributions due to varying climates. Similarly, existing approaches may fail to capture fine-grained domain-specific information from numerical data distributions. For instance, financial transaction amounts and sales figures might overlap in certain ranges but differ in others, presenting specific challenges such as variability in data distributions and similar contextual information. Numerical columns often have diverse distributions, such as normal, skewed, or multimodal, which can be challenging to model accurately. Existing methods may struggle to differentiate between columns with similar value ranges but different underlying distributions. Additionally, many approaches rely heavily on contextual information from table names and neighboring columns, which might not always be available. This reliance can lead to misclassification when context is absent or incomplete.

In this paper, we aim to address challenges associated with numerical data, and propose an approach based on a Gaussian Mixture Model (GMM) \cite{dempster1977maximum,reynolds2009gaussian,pearson1894contributions} to identify data distributions existing in different columns. Gem focuses solely on numerical columns without utilizing context from table names or neighboring columns. However, we later incorporate context from column headers (attribute names) to investigate how numeric-only embeddings contribute to improvements in downstream tasks. We defined a signature mechanism to draw a probability matrix from each column, which shows the probability of a column belonging to a particular Gaussian component or distribution, which can be interpreted as a semantic type.

The contributions of this paper are: 
\begin{enumerate}
   \item We introduce a method for producing embeddings for numerical columns that leverages GMMs to handle numerical distributions in tabular data. This approach utilizes the statistical properties of distributions and a unique signature method to form a probability matrix from Gaussian distributions, focusing exclusively on numerical data.

    \item We investigate the contribution of embeddings produced from numerical values in combination with header information. This includes thoroughly analyzing the impact of integrating numerical data distributions and header embeddings using transformer models.
    
    
    \item A comprehensive comparative analysis of Gem against state-of-the-art bespoke methods reveals that Gem consistently achieves superior performance, both when incorporating contextual information and when using only numerical values.
\end{enumerate}

\section{Related Work}

We review the literature on embeddings for numerical data in two categories: (i) approaches that employ GMMs and other mixture models, and (ii)  numerical embedding methods for tabular data. 

\subsection{Mixture models for embeddings}
Several approaches have adopted mixture models and other distributional techniques to encode numerical data via distributions. These methods have proven helpful for various downstream tasks. One notable method \cite{DBLP:conf/nips/GorishniyRB22} proposes two mechanisms to encode numerical data: piecewise linear encoding and periodic activation functions. Piecewise linear encoding divides the numerical range into segments and fits linear functions within each segment, capturing non-linear relationships among numerical features. Periodic activation functions map numerical values to a higher-dimensional space using sinusoidal transformations, which helps capture periodic patterns. Another related approach, \cite{DBLP:conf/emnlp/JiangNGCZST20}, introduces a method utilizing Self-Organizing Maps (SOM) and GMM to create numeral embeddings. Both SOM and GMM integrate numerical values and neighboring textual data to produce embeddings, which are calculated as weighted averages of prototype numeral embeddings determined by a similarity function and integrated into traditional word embeddings. Unlike the previous method \cite{DBLP:conf/nips/GorishniyRB22}, focusing specifically on either numerical data or non-numerical data, this method \cite{DBLP:conf/emnlp/JiangNGCZST20} embeds numeral contexts within traditional word embeddings, enhancing numeral understanding in text data applications. 

The MULTIHIERTT framework \cite{DBLP:conf/acl/ZhaoLLZ22} handles numerical reasoning over hybrid datasets that integrate hierarchical tables and textual data. It is developed on MT2Net \cite{DBLP:conf/acl/ZhaoLLZ22}, combining numerical data and contextual information from table structures, such as headers and neighboring text. Unlike GMM and SOM \cite{DBLP:conf/emnlp/JiangNGCZST20}, which mainly focus on numerical columns, MT2Net enhances its reasoning capabilities by utilizing the structural metadata from tables. The MULTIHIERTT framework works in two stages: a fact-retrieval module first gathers relevant numerical and textual information, followed by a reasoning module that applies both symbolic and arithmetic operations to integrate and reason over the retrieved data. Several other recent approaches adopt numerical reasoning methods to embed numerical data, each tailored to specific applications, including number decoding \cite{DBLP:conf/emnlp/WallaceWLSG19}, automatic data generation \cite{DBLP:conf/acl/GevaGB20}, numerical attribute estimation \cite{DBLP:journals/corr/abs-2109-03137,DBLP:conf/kdd/KimKKPJP23} and data-to-text generation \cite{DBLP:conf/acl/SuadaaKFOT20}.

\subsection{Numerical embeddings for tabular data}
In applications such as semantic column type detection, it is challenging to detect the type of a column solely relying on numerical data, and as a result many proposals combine numerical features with other context.
Our work aims to increase the extent to which numeric column values can inform such applications by considering distributions from the column values themselves. 

Most existing approaches consider contextual evidence from neighboring columns, rows, tables, and metadata, such as table descriptions or names. For instance, DICE \cite{DBLP:conf/emnlp/SundararamanSSW20} produces embeddings to reflect actual distances on the number line, utilizing contextual information from surrounding words to enhance numerical reasoning. This involves creating vector representations (embeddings) for numerical values such that the cosine similarity between these embeddings corresponds to the numerical difference between the values. For example, if two numbers in a column are 5 and 10, the DICE embeddings would ensure that the vector representations for these numbers are placed in such a way that the cosine similarity between them reflects the numerical distance of 5 units.

Research has often focused on leveraging as much contextual evidence as possible to improve column semantic type detection. For example, Sato \cite{DBLP:journals/pvldb/ZhangSLHDT20} uses column values from numeric and non-numeric neighboring columns, table metadata, and global table features. Sato employs a structured prediction model that integrates unary potentials (functions that represent certain relationships in the structured prediction model) from individual columns and pairwise potentials between adjacent columns to capture inter-column semantic relationships. These relationships define the embeddings within the same table, as certain semantic types often co-occur across columns. For instance, a \textit{"Date"}  column is semantically related to a \textit{"Payment Due"} column in finance data. Similarly, Sherlock \cite{DBLP:conf/kdd/HulsebosHBZSKDH19} is a multi-input neural network-based architecture that detects semantic data types by analyzing features extracted from numerical and non-numerical contexts. Sherlock utilizes metadata from column headers, adjacent textual data within the table, and numerical values. Sherlock composes features to generate comprehensive embeddings, including character distributions, word embeddings, and paragraph vectors. Unlike Sherlock, which focuses on intra-table context, RECA \cite{DBLP:journals/pvldb/SunXC23} extends the contextual scope by incorporating data from related tables, providing a broader contextual framework. RECA utilizes a graph neural network to integrate features from related tables, capturing complex inter-table relationships. In contrast, Doduo \cite{DBLP:conf/sigmod/SuharaL0ZDCT22} employs a pre-trained Transformer-based language model and multi-task learning to predict column types and relations within tables. Doduo considers non-numerical context, such as textual data from cell values, and uses a multi-column approach with attention mechanisms to capture fine-grained token-level interactions among cells within the same table. This differs from Sherlock and RECA in that they focus primarily on the inherent data within table cells rather than the external context. 

A recent proposal, Pythagoras \cite{DBLP:conf/edbt/LangeneckerSSB24}, outperforms Sato \cite{DBLP:journals/pvldb/ZhangSLHDT20}, Sherlock \cite{DBLP:conf/kdd/HulsebosHBZSKDH19}, and Doduo \cite{DBLP:conf/sigmod/SuharaL0ZDCT22} by focusing on a more holistic integration of numerical and non-numerical contexts. Pythagoras employs a Graph Neural Network (GNN) and constructs a heterogeneous graph to integrate various contextual signals, including table names, numerical data, and metadata from neighboring columns. Unlike earlier methods that either emphasize related tables (RECA \cite{DBLP:journals/pvldb/SunXC23}), intra-table relationships (Sherlock \cite{DBLP:conf/kdd/HulsebosHBZSKDH19}), or token-level interactions (Doduo \cite{DBLP:conf/sigmod/SuharaL0ZDCT22}), Pythagoras synthesizes these contexts within a unified graph structure.

Existing solutions integrate numerical and non-numerical metadata to enhance detection accuracy but do not rely exclusively on numerical data; these methods consider the additional context from neighboring columns, table metadata, and textual information. In terms of numerical embeddings, most existing approaches focus on statistical properties of numeric columns rather than distributional properties, e.g., Pythagoras \cite{DBLP:conf/edbt/LangeneckerSSB24} and Sato \cite{DBLP:journals/pvldb/ZhangSLHDT20}. Conversely, \textit{ad hoc} methods that capture distributional properties based on Gaussian Mixture Models (GMMs) or similar techniques have proven effective for other numerical tasks, such as clustering and density estimation. However, these approaches have not been widely generalized for data management tasks within tabular data.
As a result, there remains a gap in fully leveraging the numerical features of data. Existing methods do not focus on drawing distributions from numerical columns and clustering them based on similar distributions, missing the opportunity to optimize the annotations of semantic types by utilizing the inherent properties of numerical data. Our approach addresses this gap by extracting numerical data distributions. We cluster columns with similar distribution profiles, allowing for a more precise understanding and categorization of numerical data.

\section{Gem-based signatures for numerical columns}
\begin{figure*}[!htbp]
  \centering
  \includegraphics[width=1.0\linewidth,keepaspectratio]{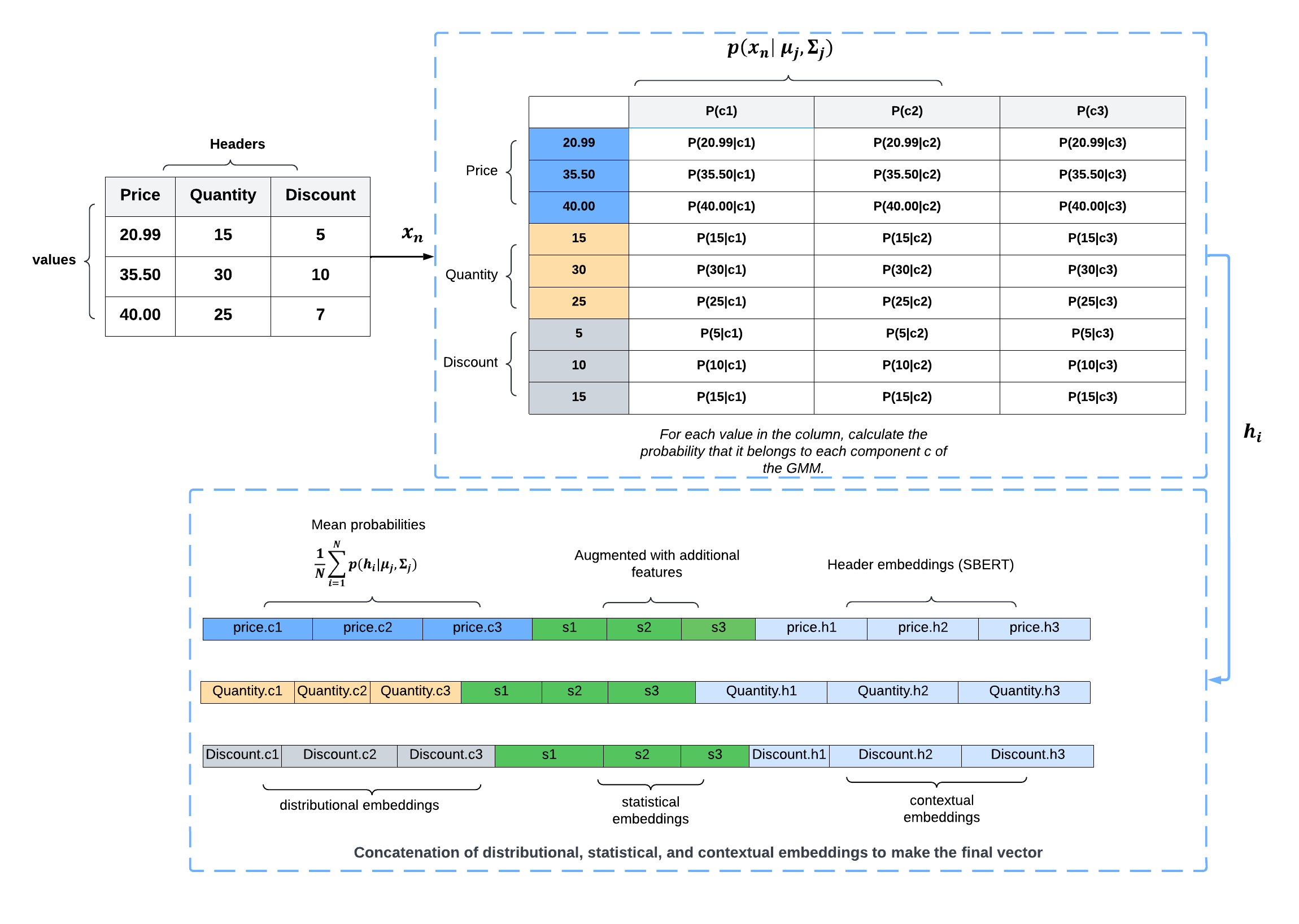}
  \caption{The process of transforming a table with three numeric columns \textit{(Price, Quantity, Discount)} into a final embedding matrix. First, the GMM is fitted to the values in each column. For each value \( x_n \) in a column, the probability \( p(x_n \mid \mu_j, \Sigma_j) \) that it belongs to each component \( C_j \) of the GMM is calculated using Equation \ref{eq6} where \( \mu_j \) and \( \Sigma_j \) are the mean and covariance matrix of component \( j \), respectively. Next, the mean probabilities for each component are computed:
$ \mu_{C_j} = \frac{1}{N} \sum_{i=1}^{N} p(h_i \mid \mu_j, \Sigma_j) $
where \( N \) is the number of values in the column. These mean probabilities are augmented with additional statistical features $(s1, s2, s3 ... sn)$. Simultaneously, the column headers are transformed into embeddings using the SBERT model. Finally, the normalized probability matrix (value embeddings) and the normalized SBERT embeddings (header embeddings) are combined to form the final embedding matrix for the table. The final embedding vector for each column includes the distributional embeddings using GMM, the statistical embeddings using data properties, and the contextual embeddings from headers, resulting in a comprehensive representation of the column data.}
  \label{fig:proposedframwork}
\end{figure*}

Building on the related work, Gem seeks to address the limitations of existing methods by maximizing the use of numerical data distributions to generate embeddings. Our proposed method uses numerical data distributions to identify and cluster columns with similar semantic types. Gem provides a unique approach to tackling numeric columns by grouping distributions (in other words, histograms) from tables that refer to the same semantic type. Additionally, it is designed to combine the numerical embeddings with other types of evidence, as explored in the experiments (see Section \ref{results}). This process involves extracting numerical data from tabular data, fitting a GMM to capture their distributional characteristics, and then calculating a probability matrix for each column based on these distributions. Gem takes stacks of numerical columns and uses a signature mechanism to predict the probabilities of each column belonging to corresponding Gaussian components. These probabilities are then aggregated for each column to form a likelihood distribution across the different components, effectively capturing the underlying numerical characteristics. Gem then calculates additional statistical features for each column and integrates contextual information from headers. These combined features enhance clustering, distinguishing similar distributions based on distributional, statistical, and contextual properties. In Figure \ref{fig:proposedframwork}, we illustrate the transformation of numerical columns into final embeddings. In the following, we describe Gem for producing embeddings from numerical columns.

\subsection{Modelling Value Distributions Using GMMs}

Assume we have a dataset comprising \( n \) columns, each representing a distinct set of numerical values. The primary representational goal is to capture the underlying distributions from which the values in each column are drawn. GMM offers an approach to this problem, leveraging the ability to deliver an expressive probabilistic model to identify the latent Gaussian distributions that collectively describe the numerical values.

A GMM is a probabilistic model representing a mixture of \( m \) Gaussian distributions. GMM represents the dataset's probability density function (pdf) as a weighted sum of multiple Gaussian distributions. The pdf of a GMM is given by \cite{reynolds2009gaussian,pearson1894contributions}:

\begin{equation}
\label{eq1}
p(x) = \sum_{j=1}^{m} \pi_j \mathcal{N}(x|\mu_j, \Sigma_j)
\end{equation}

\noindent where:
\begin{itemize}
    \item \( x \) is a numeric value.
    \item \( \pi_j \) is the mixing coefficient for the \( j \)-th Gaussian component, with \( \sum_{j=1}^{m} \pi_j = 1 \).
    \item \( \mathcal{N}(x|\mu_j, \Sigma_j) \) is the Gaussian distribution with mean \( \mu_j \) and covariance \( \Sigma_j \).
\end{itemize}

To estimate the parameters (\(\mu_j\), \(\Sigma_j\) and \(\pi_j\)) of the GMM, we employ the Expectation-Maximization (EM) algorithm \cite{dempster1977maximum}, which iteratively optimizes these parameters to maximize the likelihood of the observed numeric columns. The EM algorithm includes two main steps: the Expectation step (E-step) and the Maximization step (M-step). Initially, the means \(\mu_j\), covariances \(\Sigma_j\), and weights \(\pi_j\) are initialized randomly. In the E-step, the responsibilities \(\gamma(z_{nj})\) are calculated, which represent the probability that a data point \(x_n\) belongs to the \(j\)-th Gaussian component:

\begin{equation}
\label{eq2}
\gamma(z_{nj}) = \frac{\pi_j \mathcal{N}(x_n \mid \mu_j, \Sigma_j)}{\sum_{k=1}^{m} \pi_k \mathcal{N}(x_n \mid \mu_k, \Sigma_k)}
\end{equation}

In the M-step, the parameters are updated based on the responsibilities computed in the E-step:

\begin{equation}
\label{eq3}
\mu_j = \frac{\sum_{n=1}^{N} \gamma(z_{nj}) x_n}{\sum_{n=1}^{N} \gamma(z_{nj})}
\end{equation}

\begin{equation}
\label{eq4}
\Sigma_j = \frac{\sum_{n=1}^{N} \gamma(z_{nj}) (x_n - \mu_j)(x_n - \mu_j)^T}{\sum_{n=1}^{N} \gamma(z_{nj})}
\end{equation}

\begin{equation}
\label{eq5}
\pi_j = \frac{1}{N} \sum_{n=1}^{N} \gamma(z_{nj})
\end{equation}

This iterative process continues until convergence, typically when the change in the likelihood of the data given the parameters falls below a pre-defined threshold. In our case it is the default value of 1e-3. Once we obtain Gaussian components from the GMM for each numeric value, we use the Gem signature mechanism to compute the likelihood that each column belongs to a particular Gaussian component.

\subsection{Gem Signature Mechanism}

In this step, Gem treats all numerical values from the columns as a single stack (one-dimensional array) of numeric values rather than individual columns. Here, signatures refer to feature vectors extracted from each column, capturing essential characteristics for analysis. For each data point \(x_n\), we compute the probability of it being generated by each Gaussian component \(j\) using the fitted parameters of the GMM (see Figure \ref{fig:proposedframwork}). This is done using pdf \cite{reynolds2009gaussian,pearson1894contributions}:

\begin{equation}
\label{eq6}
p(x_n \mid \mu_j, \Sigma_j) = \frac{1}{(2\pi)^{d/2} |\Sigma_j|^{1/2}} \exp \left( -\frac{1}{2} (x_n - \mu_j)^T \Sigma_j^{-1} (x_n - \mu_j) \right)
\end{equation}

Using these probabilities, we construct a probability matrix \(P\), where each element \(P_{nj}\) represents the responsibility \(\gamma(z_{nj})\) computed during the E-step. The matrix \(P\) thus encodes the likelihood of each data point belonging to each Gaussian component, effectively summarizing the distributional characteristics captured by the GMM. 

In addition to these GMM-derived probabilities, we extract several statistical features from each numeric column to capture the statistical aspects of the column's distribution. These features were selected by systematically evaluating the Pythagoras feature set \cite{DBLP:conf/edbt/LangeneckerSSB24}. We only focus on features applicable to numerical values. Each feature's correlation with the Gann embeddings was tested, and only those with high correlation were retained. The select features includes:
\begin{itemize}
    
\item \textit{Unique count}:  Reflects the variety of distinct values in the column, indicating whether the data is largely repeated.
\item \textit{Mean}: Representing the average value in the column.
\item \textit{Coefficient of variation (CV)}: A normalized measure of spread that indicates the relative dispersion of values.
\item  \textit{Entropy}: Quantifies the degree of uncertainty in the data distribution.
\item  \textit{Range}: The difference between the maximum and minimum values.
\item  \textit{Percentiles (10th and 90th)}: Highlight the lower and upper bounds to provide insights into the data's distribution.
\end{itemize}

Mathematically, let \( \mathbf{f}_i \) represent the vector of additional features for the \( i \)-th column. Standardization transforms each feature vector \( \mathbf{f}_i \) to \( \mathbf{\tilde{f}}_i \), where:

\begin{equation}
\label{eq7}
\mathbf{\tilde{f}}_i = \frac{\mathbf{f}_i - \mu(\mathbf{f})}{\sigma(\mathbf{f})}
\end{equation}

where \( \mu(\mathbf{f}) \) and \( \sigma(\mathbf{f}) \) are the mean and standard deviation of the feature vector, respectively. These standardized feature vectors are then integrated with the mean probabilities derived from the GMM. For the \( i \)-th column, let \( \mathbf{m}_i \) represent the mean probability vector of length \( K \) (number of Gaussian components). The augmented feature vector \( \mathbf{a}_i \) is formed by concatenating \( \mathbf{m}_i \) and \( \mathbf{\tilde{f}}_i \):

\begin{equation}
\label{eq8}
\mathbf{a}_i = [\mathbf{m}_i \, \| \, \mathbf{\tilde{f}}_i]
\end{equation}

Finally, each augmented feature vector \( \mathbf{a}_i \) is normalized to ensure comparability across different columns, resulting in the final row of the probability matrix \( P_i \):

\begin{equation}
\label{eq9}
P_i = \frac{\mathbf{a}_i}{\|\mathbf{a}_i\|_1}
\end{equation}

where \( \|\mathbf{a}_i\|_1 \) denotes the L1 norm of \( \mathbf{a}_i \). This integration enhances the descriptive power of the probability matrix by combining the probabilistic information (distributional embeddings) from the GMM with additional statistical characteristics of the columns.

\subsection{Header embeddings}

We obtain the distributional embeddings from the GMM and capture the statistical properties from the column values. Additionally, we incorporate contextual information from the headers. This step, while not always necessary, proves helpful when the distributional embeddings are highly dense and similar. The context provided by the column headers helps to disambiguate meaningful differences among columns. Our experiments (see Section \ref{results}) report results with and without headers, demonstrating Gem's flexibility and effectiveness. We use Sentence-BERT (SBERT) \cite{DBLP:conf/emnlp/ReimersG19} to embed column headers, which captures the semantic meaning of the headers in a high-dimensional space. Let \( \mathbf{s}_i \) represent the SBERT embedding for the \( i \)-th column header. To ensure compatibility with the value embeddings, the SBERT embeddings are also normalized:

\begin{equation}
\label{eq10}
\mathbf{S}_i = \frac{\mathbf{s}_i}{\|\mathbf{s}_i\|_1}
\end{equation}

where \( \|\mathbf{s}_i\|_1 \) denotes the L1 norm of \( \mathbf{s}_i \).

Finally, the normalized probability matrix \( \mathbf{P}_i \) (value embeddings) and the normalized SBERT embeddings \( \mathbf{S}_i \) (header embeddings) are concatenated to form the final combined embedding vector \( \mathbf{C}_i \) for each column:

\begin{equation}
\label{eq11}
\mathbf{C}_i = \left[ \mathbf{P}_i \| \mathbf{S}_i \right]
\end{equation}

where \( \| \) denotes the concatenation operation. This final embedding vector \( \mathbf{C}_i \) includes the probabilistic and semantic information, providing a joint representation of each column in the table for downstream tasks (attribute name + represented value distribution). For example, in clustering, each data point is assigned to the Gaussian component with the highest responsibility:

\begin{equation}
\label{eq12}
\text{Cluster}(x_n) = \arg \max_j C_{nj}
\end{equation}

In addition to the contextual embeddings and probabilistic representations, we also aggregate standardized statistical features \( \mathbf{\tilde{f}}_i \). These standardized features \( \mathbf{\tilde{f}}_i \) are integrated with the probabilistic embeddings \( \mathbf{P}_i \) and the normalized SBERT embeddings \( \mathbf{S}_i \), forming the final aggregated embedding vector \( \mathbf{C}_i^{\text{final}} \):

\begin{equation}
\label{eq14}
\mathbf{C}_i^{\text{agg}} = \left[ \mathbf{P}_i \| \mathbf{S}_i \| \mathbf{\tilde{f}}_i \right]
\end{equation}

This final aggregated embedding \( \mathbf{C}_i^{\text{agg}} \) combines probabilistic, semantic, and standardized statistical information, providing a rich representation for downstream tasks.

By integrating the GMM with extracted signatures and leveraging the resulting probability matrix, we establish a comprehensive framework for managing and analyzing datasets composed of numerical columns. 

We provide pseudocode to generate the final embedding matrix from numerical columns in Algorithm \ref{algorithm1}.

\begin{algorithm}[!htbp]

\caption{Generating the final embedding matrix from numeric columns}
\begin{algorithmic}[1]
\label{algorithm1}
\REQUIRE Dataset with $n$ numeric columns $\{\mathbf{x}_1, \mathbf{x}_2, \ldots, \mathbf{x}_n\}$ and headers $\{h_1, h_2, \ldots, h_n\}$
\ENSURE Final embedding matrix $\{\mathbf{C}_1, \mathbf{C}_2, \ldots, \mathbf{C}_n\}$

\STATE Initialize lists: $\text{column\_headers} \leftarrow []$, $\text{additional\_features} \leftarrow []$

\FOR{each column $i$ in the dataset}
    \STATE Extract column values $\mathbf{x}_i$ and header $h_i$
    \STATE Append $h_i$ to $\text{column\_headers}$
    \STATE Calculate additional statistical features $\mathbf{f}_i$
    \STATE $\mathbf{f}_i \leftarrow [f_{1_i}, f_{2_i}, \ldots, f_{m_i}]$
    \STATE Append $\mathbf{f}_i$ to $\text{additional\_features}$
\ENDFOR

\STATE Fit GMM with $m$ components to all column values: $\text{GMM\_model} \leftarrow \text{GMM}(m)$

\FOR{each column $i$}
    \FOR{each data point $x_{i,j}$ in $\mathbf{x}_i$}
        \STATE Compute the probability $p(x_{i,j} \mid \mu_k, \Sigma_k)$ using Equation \ref{eq6}
    \ENDFOR
    \STATE Compute the mean probabilities for each component
    \STATE Form the augmented feature vector $\mathbf{a}_i$ using Equation \ref{eq8}:
    \STATE Normalize $\mathbf{a}_i$ to form $P_i$ using Equation \ref{eq9}:
    \STATE Append $P_i$ to the probability matrix
\ENDFOR

\STATE Encode column headers using SBERT: $\mathbf{s}_i \leftarrow \text{SBERT\_model.encode}(h_i)$
\STATE Normalize SBERT embeddings using Equation \ref{eq10}:

\FOR{each column $i$}
    \STATE Concatenate $P_i$ with $\mathbf{S}_i$ to form the final combined embedding vector $\mathbf{C}_i$ using Equation \ref{eq11}:
    \STATE Append $\mathbf{C}_i$ to the final embedding matrix
\ENDFOR
\STATE Output the final embedding matrix

\end{algorithmic}
\end{algorithm}

\section{Evaluation}

\begin{table*}[t]
	\centering
	\caption{Dataset statistics related to the number of numeric columns and GT clusters. The numbers in brackets indicate the columns and semantic types in the ground truth (GT), which were derived by refining coarse-grained annotations into fine-grained ones for both the GDS and WDC datasets}
	\label{tab:1}
	\begin{tabular}{ccccc}  
			\toprule
		& GDS & WDC & Sato Tables & Git Tables \\ 
		\midrule 
		\# Columns & 2491 (2117) & 2852 (5678) & 2231 & 459 \\  
		\#GT clusters & 86 (96) & 147 (325) & 12 & 19 \\ 
		\bottomrule 
	\end{tabular}
\end{table*}

\subsection{Datasets}
We use four widely used datasets, which include Sato Tables \cite{DBLP:journals/pvldb/ZhangSLHDT20}, Git Tables \cite{DBLP:journals/pacmmod/HulsebosDG23}, Google Dataset Search (GDS)\footnote{\url{https://github.com/PierreWoL/SILM}}, and Web Data Commons (WDC)\footnotemark[\value{footnote}] to evaluate Gem \footnote{\url{https://github.com/hafizrauf/Gem}}. We select numeric columns from all four datasets. The chosen datasets have been selected for their abundance of numeric columns, rich variability in data distributions, and diverse column semantics. 
Dataset details are given below and in Table \ref{tab:1}.

\begin{itemize}

 \item  \textbf{Sato Tables}, part of the VizNet dataset, includes various numeric columns representing attributes such as population counts, GDP values, and personal statistics. Many numeric columns have similar distributional characteristics but different semantic types. For instance, columns labeled as "\textit{age}," "\textit{duration}," "\textit{weight}," "\textit{order}," and "\textit{position}" exhibit similar numeric distributions, yet they have different semantic meanings. The distributional and statistical similarity between these columns is greater than 0.90, indicating their contextual (header embeddings) meanings differ significantly despite their numeric resemblance. The distributional and statistical similarity of the columns is derived using our Gem embedding method.

 \item \textbf{Git Tables} is a large-scale semantic type detection dataset consisting of relational tabular data from a wide range of domains. The column annotations were obtained from Schema.org and DBpedia. Git Tables represents a particularly challenging setting without additional context descriptions. For example, detecting the semantic type of a column given the values \textit{[153, 228, 125, 273, 319, 139, ...]} to be \textit{duration, height, length} or \textit{volume}.

 \item \textbf{WDC} (Web Data Columns) includes numeric columns extracted from web data, such as product prices, stock quantities, and review scores. It captures a broad spectrum of e-commerce and social media numeric data. WDC attribute names are categorically coarse-grained. For example, columns like \textit{Score\_Cricket, Score\_Rugby, Score\_Football} are semantically annotated with \textit{Score}. However, we transform the annotation from coarse-grained to fine-grained to better capture the different distributions of each column. For instance, while both \textit{Score\_Cricket} and \textit{Score\_Rugby} represent game scores, they have distinct contexts and distributions\_Cricket scores tend to be much higher due to the nature of the game, while Rugby scores follow a different scale. Simply classifying them as \textit{Score} would overlook these differences. Further details of the column annotation process are provided in Section \ref{annotation}.

 \item \textbf{GDS} (Google Dataset Search) is a platform developed to help researchers discover openly available datasets on the web. We used the GDS dataset, where the authors manually curated specific tables for data discovery tasks.. This dataset has been refined to a fine-grained level from its original form, ensuring that each table represents distinct and specific concepts for more precise column annotation. For example, instead of having a general \textit{"power"} column, we annotate columns with more granularity, such as \textit{"engine\_power\_car"} and \textit{"battery\_power\_device"}, which capture contextually relevant information about the power of car engines and electronic devices, respectively.
\end{itemize}

To determine the suitability of the four datasets (GDS, WDC, Sato Tables, and Git Tables) to adopt for the application of creating embeddings for numerical data, the following criteria were established:

\begin{itemize}

\item  \textbf{Numerical Columns Specificity:} Each dataset contains a significant number of columns that are composed entirely of numerical data (see Table 4). This is essential for ensuring the effectiveness of embeddings in the context of numerical columns. 

\item  \textbf{GT clusters with detailed refinement}: Another criterion is the availability of GT clusters that categorize different semantic types. The initial annotations for the GDS and WDC datasets were refined from broader, coarse-grained types to more specific, fine-grained semantic categories. For instance, the GDS dataset refined clusters from 86 to 96 distinct types, while WDC refined 147 clusters into 325 semantic types. 

\item  \textbf{Diversity across datasets:} The selected datasets provide a broad spectrum of semantic types representing different domains. While GDS and WDC are more extensive and varied datasets with a wider range of semantic types, Sato Tables and Git Tables offer smaller, more specialized datasets with 12 and 19 clusters, respectively. 

\end{itemize}

\subsubsection{Data Annotation: From Coarse-Grained to Fine-Grained Labels}

\label{annotation}
We use the following criteria to convert coarse-grained labels into fine-grained labels for both WDC and GDS datasets, as both datasets often have coarse-grained annotations as ground truth. For example, the score of a \textit{cricket} and the score of a \textit{football} game can be classified under the supertype "\textit{score}". However, semantically, they represent different meanings in the real world with different numeric distributions. Our criteria are as follows:
\begin{itemize}

\item Two columns should have the same annotation if they describe the same domain. Applying the equality $(=)$ operator to values from different columns should be meaningful. For example, it is not meaningful to compare a \textit{volume} with an \textit{area} as they have different units.

\item Two values must describe the same real-world concept for them to be equivalent. For instance, a '\textit{height}' cannot be equivalent to a '\textit{length},' as they measure fundamentally different properties despite sharing the same unit.
   
\item  If subcategories exist, they must be applied at the appropriate level of specificity. For example, the \textit{score} of a \textit{soccer game} cannot be equated to the \textit{score} of a \textit{cricket match}, even though both fall under the super category of \textit{score}. 
\end{itemize}

\subsubsection{Evaluation Metrics}
We evaluated Gem for two downstream tasks: semantic table annotation on numeric columns and column clustering. We used Precision and Recall at \textit{k} for column semantic type detection as evaluation metrics. Precision measures the fraction of the top \(k\) columns selected by the model that are relevant to a given column. Recall identifies the fraction of the relevant columns within the top \(k\) selections. The top \(k\) are the \(k\) nearest neighbors to any single column, where \(k\) equals the total number of columns with the same semantic type in the Ground Truth. We use the following measures to evaluate correct and incorrect classifications:

\begin{itemize}
    \item \textbf{\textit{True Positives (TP)}}: The columns among the top \(k\) that have the same label as the selected column.
    \item \textbf{\textit{False Negatives (FN)}}: The relevant columns (those sharing the same label as the selected column) that were not included in the top \(k\) selections.
    \item \textbf{\textit{False Positives (FP)}}: The other columns in the top \(k\) that do not share the same label.
\end{itemize}

To determine the top \(k\) neighbors, we calculate the cosine similarity matrix for all columns, and for each selected column, we sort the columns by their similarity scores in descending order. We select the top \(k\) indices, excluding the column itself, to find the \(k\) nearest neighbors.

We used two well-known evaluation metrics for column clustering: Accuracy (ACC) \cite{DBLP:journals/tip/YangXNYZ10} and Adjusted Rand Index (ARI) \cite{DBLP:conf/iccv/WuLWQLLZ19}. ACC measures the proportion of correctly clustered columns and ranges from 0 to 1. The ARI score ranges from -1 to 1, where negative values suggest worse-than-random labeling, 0 indicates random labeling, and 1 indicates a perfect match.

\subsubsection{Baselines}
Based on two main criteria, we selected the following baseline methods to compare with Gem for numerical embeddings. First, we chose methods that do not consider contextual information and rely solely on numerical data, making them reasonable candidates for comparison. Second, we included Pythagoras, Sato, and Sherlock, which typically incorporate context from headers, table names, and neighboring columns. We re-implemented Pythagoras, Sato, and Sherlock to retain their core statistical features and header information to ensure a fair comparison with our context-independent approach. However, we excluded other contexts, such as table names and neighboring columns. 

We acknowledge that this re-implementation of Pythagoras, Sato, and Sherlock is a simplified version of the original methods. However, it illustrates the specific impact of removing certain contextual elements, allowing for a more precise comparison with Gem's context-independent approach.

\begin{itemize}
    \item \textbf{\textit{Piece-wise Linear Encoding (PLE)}} \cite{DBLP:conf/nips/GorishniyRB22} transforms numeric data into a series of linear segments, each representing a portion of the data range. This method simplifies complex non-linear relationships into manageable linear parts by dividing the numeric range into intervals and applying linear transformations within each segment.

    \item \textbf{\textit{Periodic Activation Functions (PAF)}} \cite{DBLP:conf/nips/GorishniyRB22} introduce oscillatory behavior into neural network layers, making them adept at capturing repeating patterns in numeric data. This model with periodic function efficiently learns and represents cyclical patterns and can detect semantic types that exhibit periodic behavior.

    \item \textit{\textbf{Squashing\_GMM }}\cite{DBLP:conf/emnlp/JiangNGCZST20}: This method begins by squashing numeric values into log space following a prototype induction using GMM to identify the clusters, each representing a prototype. Similarity functions then measure how closely numeric columns match these Gaussian components. 
    
    \item \textbf{\textit{Squashing\_SOM}}~\cite{DBLP:conf/emnlp/JiangNGCZST20}:  This method is similar to the one above, Squashing\_GMM, except for the prototype induction part, where SOM projects the log-transformed data onto a lower-dimensional grid while preserving its topological structure, inducing prototypes representing data clusters.
\item \textbf{Kolmogorov-Smirnov (KS) statistic} \cite{massey1951kolmogorov}: We include the KS statistic as a baseline to compare with Gem. The KS statistic is particularly relevant in this context because it measures the maximum difference between the cumulative distribution functions (CDFs) of the empirical data and several theoretical distributions, such as normal \cite{fischer2011history}, uniform \cite{feller1991introduction}, exponential \cite{balakrishnan2019exponential}, beta \cite{johnson1995continuous}, gamma \cite{hogg2013introduction}, lognormal \cite{limpert2001log}, and logistic \cite{johnson1995continuous}. We evaluate how well the numerical data in columns aligns with these reference distributions; we generate features that capture the underlying semantic type of the columns because different semantic types exhibit unique distributional patterns, and the KS statistic helps identify these patterns accurately.

\item \textbf{\textit{Sherlock }}\cite{DBLP:conf/kdd/HulsebosHBZSKDH19}: We compare Sherlock with Gem because it extracts statistical features from numerical columns, such as mean, variance, skewness, and kurtosis, which align with Gem's focus on numerical data. To ensure a fair comparison, we augment these statistical features with SBERT-generated embeddings from column headers, similar to Gem's use of header information. Sherlock's model processes these combined features using dense layers with dropout and a softmax layer.

\item \textbf{\textit{Sato}}  \cite{DBLP:journals/pvldb/ZhangSLHDT20}: We also compare Sato, which is an enhancement in Sherlock. To maintain fairness, we exclude Sato's global and local context features, which rely on neighboring nonnumerical columns, since Gem does not utilize the nonnumerical global context. In our implementation of Sato, we focus on single-column data, extracting the same statistical features as Sherlock and combining them with SBERT embeddings from the headers. These combined features are processed in Sato's neural network model. Overall, we extract statistical features in both implementations (Sherlock and Sato) and combine them with SBERT embeddings before processing them through their respective training architectures to obtain embeddings.

\item \textbf{\textit{Pythagoras}} \cite{DBLP:conf/edbt/LangeneckerSSB24}: Pythagoras uses a graph representation of tables to capture both numerical and contextual information, such as table names and neighboring columns. The model combines pre-trained language models for initial encoding with specialized subnetworks for numerical features. In line with Gem’s focus on numerical data and headers only, we re-implemented Pythagoras in a context-reduced version, where only header data was considered, excluding table names and neighboring columns. Additionally, we retained the same statistical features selected for Gem.

\end{itemize}

To differentiate the original versions of Pythagoras, Sherlock, and Sato from our adapted versions, we called them Pythagoras\_{SC}, Sherlock\_{SC}, and Sato\_{SC}, where SC indicates \textit{Single-Column}. This shows that our implementations work with individual numerical columns without relying on multi-column and table-wide context. In the original approaches, these methods use additional information, such as neighboring columns and metadata. However, in our adapted versions, we remove this extra context to focus exclusively on the features of single-column numerical data to ensure the comparison with Gem is fair and consistent.

\subsubsection{Parameter Setting}
The number of Gaussian components does not significantly impact Gem's overall performance (see ablation study in Section \ref{ablation_components}). Through comprehensive experimentation, we found that each column generally exhibits between 5 to 10 distinct distributions, and further increasing the number of components beyond this range does not contribute to performance improvement. Specifically, using more than 10 Gaussian components per column leads to model complexity without corresponding gains in accuracy. However, we determine each dataset's optimal number of components using the Bayesian Information Criterion (BIC). The BIC results showed consistent performance across 5 to 100 components, with minimal fluctuations. To maintain consistency, we used 50 Gaussian components for all our analyses. In baselines, specifically Squashing\_GMM \cite{DBLP:conf/emnlp/JiangNGCZST20}, we use the same number of components as used in Gem; in Squashing\_SOM \cite{DBLP:conf/emnlp/JiangNGCZST20}, PLE \cite{DBLP:conf/nips/GorishniyRB22} and PAF \cite{DBLP:conf/nips/GorishniyRB22}, we use 50 prototypes, bins, and frequencies, respectively. Additionally, we initialize the EM algorithm 10 times to increase the likelihood of finding the global optimum, ensuring robust convergence and avoiding local minima.

\begin{table}[t]
	\centering
	\caption{\textbf{Average Precision Score on GitTables, Sato Tables, GDS, and WDC datasets. We focused exclusively on numeric-only data. To ensure a consistent comparison, we used the coarse-grained versions of the GDS and WDC datasets, as GitTables and Sato Tables lack fine-grained details. Gem (D+S) represents the Distributional and Statistical components of Gem.}}
	\label{tab:2}
	\begin{tabular}{ccccc} 
		\toprule
		& \textbf{Git} & \textbf{Sato} & \textbf{WDC} & \textbf{GDS} \\ 
  		& \textbf{Tables} & \textbf{Tables} & & \\ 
		\midrule
		Squashing\_GMM\cite{DBLP:conf/emnlp/JiangNGCZST20} & 0.25 & 0.28   &0.18  & 0.29\\  
		Squashing\_SOM \cite{DBLP:conf/emnlp/JiangNGCZST20} & 0.19 & 0.31  & 0.14 & 0.28\\
		PLE \cite{DBLP:conf/nips/GorishniyRB22} & 0.19 & 0.11  & 0.18 & 0.11\\
		PAF \cite{DBLP:conf/nips/GorishniyRB22} & 0.24 &0.23 & 0.17  & 0.34\\
        KS statistic \cite{massey1951kolmogorov}  & 0.21 & 0.21 & 0.02 & 0.21\\
		Gem (D+S) & \textbf{0.28} & \textbf{0.37}  & \textbf{0.21} &\textbf{ 0.37}\\
		\bottomrule
	\end{tabular}
\end{table}

\subsection{Results and Discussion}
\label{results}
\subsubsection{Numeric-Only Results}

Table \ref{tab:2} shows the experimental results of Gem compared to the baselines, considering numeric-only data across all four datasets. We observe the following:
\begin{enumerate}

\item  \textbf{Gem consistently outperforms the baseline methods when considering numeric columns, achieving the highest average precision in all datasets}. Notable improvements relative to the best baseline are in Sato Tables (0.06), Git Tables (0.03), and GDS (0.03), demonstrating its ability to handle diverse numeric data distributions. 

\item \textbf{Baseline methods, including PLE, PAF and the KS statistic, struggled to differentiate between columns with superficially similar value ranges across all datasets.} For example, the columns labeled \textit{'Rating'} \textit{[3.6, 3.8, 3.9, 3.9, 3.6, ...]} and \textit{'Weight'} \textit{[1.0, 1.0, 1.4286, 1.25, 1.0957, 2.5, ...]} were incorrectly identified as highly similar (as evidenced by high cosine similarity in the embeddings produced by PLE and PAF, and low KS statistic values), despite having distinct value distributions and underlying semantics. In contrast, Gem distinguished between these columns, correctly classifying them as true negatives. This demonstrates Gem's superior ability to capture and identify semantic differences based on the underlying value distributions of the columns.

\item \textbf{Gem better accounts for distributional variations in detecting column semantic types.} For instance, Gem correctly identifies (true positives) the top 10 neighbors of the column '\textit{Mileage'} with values \textit{[5, 117000, 92000, 500...]} as \textit{'Mileage'} on GDS. However, with Squashing\_GMM and KS statistic, the top 10 neighbors are columns about \textit{'Rank'} and \textit{'Year'} due to the overlap in value ranges, even though these columns represent different domains. The additional statistical features combined with distributional properties in Gem effectively identify the fine-grained components among numeric columns. 

\item  \textbf{Gem accurately distinguishes between \textit{width} and \textit{length} columns in contrast with Squashing\_SOM and Squashing\_GMM using the Git Tables dataset.} For example, for the column \textit{'width'} \textit{[5, 256, 5, 256, 5.12]}, Gem achieves a precision of 0.61 compared to 0.41 with Squashing\_SOM and 0.39 with Squashing\_GMM. Squashing\_SOM and Squashing\_GMM methods incorrectly grouped \textit{'width'} together with \textit{'length'} \textit{[256, 5, 256, 5, 256, 109.71, 51.2]}, whereas Gem accurately distinguishes these columns. 

\item  \textbf{For Sato Tables, Gem tends to misclassify columns with similar value distributions, resulting in overlapping errors.} For instance, \textit{'weight'} columns with values \textit{[32.2, 34.3]} were consistently misclassified as \textit{'age' } due to their repetitive values (e.g., \textit{[32, 30, 30, 31, 31, 31, 30, 31, 31, 31]}). Conversely, despite having a low overlap in value similarities, PLE and PAF methods still resulted in mis-classifications. For example, \textit{'year'} columns with values ranging from \textit{[1980, 1981, 1982, ..., 2012]} were misclassified as \textit{'duration'} with non-similar values like \textit{[214.0, 306.0, 248.0, ...]} or \textit{'age'} with values \textit{[24, 38, 36, ...]}. This indicates that PLE and PAF struggle to differentiate columns with distinct value ranges due to insufficient semantic differentiation. In contrast, Gem, which effectively handles distributional properties, achieves more accurate classification by leveraging detailed distributional and contextual features. 

\item \textbf{Compared to PLE and PAF, Gem successfully differentiates between columns with overlapping numerical ranges by learning distributional embeddings and capturing statistical features from Sato Tables.} For instance, two columns—one representing \textit{'weight'} with values [32.2, 34.3] and another representing \textit{'age'} with values [30, 31, 34]. PLE and PAF struggle with these because of the overlap in numerical values. However, Gem effectively determines the difference by identifying that \textit{'weight'} values follow a continuous distribution, while \textit{'age'} exhibits a clustered distribution at specific points.

 \item 
\textbf{Gem consistently maintains high similarity (semantic similarly of embedding vectors) scores even when columns with the same semantic types have varying cardinalities, outperforming PAF and KS statistics}. For instance, Gem analyzes a column \textit{'year'} with 33 distinct values against another year column with 48 distinct values. Despite the difference in cardinality, Gem put them in a single cluster compared to PAF and KS statistics, which classify them into different clusters.

\end{enumerate}

\begin{table}[!t]
	\centering
\caption{Average precision score considering headers + values on fine-grained versions of GDS and WDC. In Gem, D represents Distributional data, S represents statistical data and C represents Contextual data.} 
	\label{tab:3}
	\begin{tabular}{ccc} 
		\toprule
		&  \textbf{WDC} & \textbf{GDS} \\ 
		\midrule
		SBERT (headers only)&  0.37& 0.79\\
   Pythagoras\_{SC} \cite{DBLP:conf/edbt/LangeneckerSSB24} &  0.02 & 0.01\\
   Sherlock\_{SC} \cite{DBLP:conf/kdd/HulsebosHBZSKDH19} &  0.002 & 0.27\\
   Sato\_{SC} \cite{DBLP:journals/pvldb/ZhangSLHDT20} &  0.003 & 0.25\\
		Gem (D+S) &  0.14 & 0.45\\
  Gem D+S+C (aggregation)&  0.41& 0.81\\
  Gem D+S+C (AE)&  0.40& 0.81\\
  Gem D+S+C (concatenation)&  \textbf{0.43}& \textbf{0.82}\\
		\bottomrule
	\end{tabular}
\end{table}

\subsubsection{Numeric + Headers Results}

In this section, we examine if the numerical embeddings obtained using Gem can contribute to further improvements when considering more evidence from columns to detect the semantic types. To achieve this, we obtained header embeddings for two datasets, GDS and WDC, and composed them with value embeddings using different composition approaches. 

In Gem, we experimented with three composition methods to merge embeddings: concatenation, aggregation, and learning embeddings through autoencoders (AE). In the concatenation approach, the probabilistic features from the GMM, statistical features from the columns, and contextual embeddings from the headers are combined into a single vector by joining them side by side. In contrast, the aggregation approach summarizes these different embeddings into a single representation. The third approach, learning embeddings through autoencoders, compresses the combined information into a lower-dimensional latent space.  We record the average precision score for individual headers-only and their composed versions in Table \ref{tab:3}. 

For this experiment, we use the fine-grained versions of both datasets. We observe the following: 

\begin{enumerate}

 \item \textbf{GDS headers are more semantically and syntactically distinct, making semantic type differentiation easier}. For instance, GDS headers like \textit{"age"} and \textit{"height"} clearly represent different semantic types and have distinct syntactic structures. As a result, SBERT achieved an average precision of 0.79 using GDS headers. Conversely, WDC headers are more complex and often overlap semantically and syntactically. For example, headers like \textit{"rating"} could apply to various domains such as books, movies, or hotels, leading to SBERT obtaining only 0.37 average precision with WDC headers. This highlights the challenge of high semantic and syntactic overlap in WDC. 
 
 \item  \textbf{Gem improves the average precision relative to SBERT (headers only) by 0.06 on WDC and 0.03 on GDS when we compose value embeddings with header embeddings through concatenation}. As we calculate precision for each semantic type and then aggregate all the precisions, a higher average precision reflects consistently better performance across multiple semantic types rather than isolated success in a few semantic types.
 
  \item  \textbf{Concatenation proved to be the most effective composition method for both datasets compared to aggregation and learning embeddings through AE when header contextual embeddings are combined with value embeddings}. The concatenation method preserves the integrity of each embedding type, ensuring that distributional features, statistical characteristics, and semantic context from the headers are all maintained in the final embedding. This allows the model to leverage both numerical properties and semantic signals effectively. At the same time, aggregation with three embeddings into a single representation risks losing some information as it compresses diverse characteristics into a less detailed form. In the case of AE, it is effective for capturing high-level patterns and lacks in capturing specific details, particularly in scenarios where distributional properties are less distinct. This results in a loss of granularity crucial for tasks requiring precise differentiation between similar value distributions. 
 \item \textbf{Gem's distributional embeddings help to improve the classification of overlapping header-only embeddings}. For example, in the WDC dataset, columns such as \textit{'Rating\_Movie' [10, 10, ...10]}, \textit{'Rating\_Book [5, 3, 5 ...5]},' and \textit{'Rating\_Hotel [4.0, 5.0, 0.0, 3.0, 5.0, 0.0, 4.0, ..., 5.0, 5.0, 3.0]}' are clustered together using SBERT due to their high syntactic similarity. However, while all three columns represent ratings on a 1-10 scale, Gem's distributional embeddings capture the different rating patterns within each column. For example, \textit{'Rating\_Movie'} shows a constant pattern, \textit{'Rating\_Book'} exhibits moderate variation, and \textit{'Rating\_Hotel'} includes a wider spread with lower and zero ratings. This demonstrates how integrating distributional data with contextual embeddings enhances the accuracy of embeddings. 
 
 \item \textbf{Pythagoras\_{SC}, which relies solely on headers as context for numerical embeddings, has demonstrated significant limitations when applied to the GDS dataset, where the headers are highly diverse.} It struggles to distinguish between columns with similar values, even when the headers differ. For instance, the column \textit{"Acceleration"} was incorrectly identified as being highly similar to columns like \textit{"Age"} and \textit{"Dry weight"}. Pythagoras\_{SC}'s dependence on header context proved insufficient, whereas Gem performed better in these scenarios. Likewise, on the WDC dataset, where the header information is more complex and heterogeneous, Pythagoras\_{SC} produced poorer results, as its GCN model failed to combine contextual and statistical features effectively.

 \item \textbf{GMM significantly underpins the performance of Gem by identifying distinct data distributions within columns}. For example, in column \textit{'Height\_Mountain'}, when using headers only, the precision was 0.3736, with 114 false positives. Gem improved the precision to 0.4835, with the number of false positives reduced to 47. Similarly, in column \textit{'Market\-Value}', with headers only, the precision was  0.83, with ten false positives. Gem increased the precision to 0.94, reducing false positives to 1.

\item \textbf{Sherlock\_{SC}, which relies on statistical features and embeddings derived solely from headers, shows a substantial difference in performance across the two datasets.} On the WDC dataset, where header information is more complex and varied, the precision drops significantly to 0.002, suggesting that Sherlock\_{SC} struggles to generalize across more diverse column types. However, on the GDS dataset, where the column headers are more standardized, Sherlock\_{SC} performs better with a precision of 0.27; however, in both cases, it is outperformed by Gem. For example, one mis-classification can be seen when Sherlock\_{SC} embedding vectors of two columns have a high similarity score of 0.99: one containing years of publication for books \textit{([2019, 1990, 2019, 2018])} and another with telephone data related to hotels  \textit{([13.943, 13.837])}, confusing these distinct categories of \textit{"Book"} and \textit{"Hotel"}.

\item \textbf{Similar to Sherlock\_{SC} and Pythagoras\_{SC}, Sato\_{SC}, which also uses header-based embeddings, demonstrates difficulties on the WDC dataset, achieving a precision of only 0.003.} In contrast, on the GDS dataset, Sato\_{SC} performs worse than Sherlock\_{SC} and Gem but better than Pythagoras\_{SC}, with a precision of 0.25. For example, the embedding vectors of two columns using Gem embeddings show a high similarity score of 0.98: one column contains house prices in various cities \textit{([320000, 450000, 210000])}, while the other represents population sizes in different regions \textit{([50000, 120000, 30000])}. Despite their distinct semantic categories of \textit{"Economic"} and \textit{"Demographic"}, Gem embeddings were unable to distinguish between the two.

\end{enumerate}

\subsection{Ablation Study}
\label{ablation}

We conducted an ablation study to understand the contribution of each feature type in Gem to numerical embeddings' performance. We tested combinations of Gem's distributional, statistical, and contextual feature types and calculated the average precision for each semantic type across the WDC and GDS datasets. The feature combinations we evaluated were Distributional (D), Statistical (S), Contextual (C), Distributional + Statistical (D+S), Contextual + Statistical (C+S), Distributional + Contextual (D+C), and Distributional + Contextual + Statistical (D+C+S).  

\begin{figure}[t]
\centering
\begin{tikzpicture}
    \begin{axis}[
        ybar,
        bar width=.40cm, 
        width=0.50\textwidth, 
        height=0.35\textwidth,
        enlarge x limits=0.1, 
        ylabel={\textbf{Average Precision}},
        symbolic x coords={D, S, C, D+S, C+S, D+C, D+C+S}, 
        xtick=data,
        xticklabel style={rotate=30, anchor=north east, font=\bfseries},
        ymin=0, ymax=0.9, 
        ytick={0,0.1,0.2,0.3,0.4,0.5,0.6,0.7,0.8,0.9}, 
        nodes near coords, 
        every node near coord/.append style={font=\scriptsize, yshift=0.15cm}, 
        legend style={at={(0.5,1.05)}, anchor=south, legend columns=2, draw=none, font=\bfseries},
        axis lines*=left,
        axis on top,
        tickwidth=0pt,
        ylabel near ticks,
        xlabel near ticks,
        yticklabel style={/pgf/number format/fixed, font=\bfseries},
        major grid style={line width=.2pt,draw=gray!30}, 
        grid=major, 
    ]

    \addplot[fill=wdc_color, bar shift=-0.15cm] coordinates {
        (D, 0.02)
        (S, 0.14)
        (C, 0.37)
        (D+S, 0.15)
        (C+S, 0.11)
        (D+C, 0.40)
        (D+C+S, 0.43)
    };

    \addplot[fill=gds_color, bar shift=0.15cm] coordinates {
        (D, 0.30)
        (S, 0.39)
        (C, 0.79)
        (D+S, 0.45)
        (C+S, 0.40)
        (D+C, 0.81)
        (D+C+S, 0.82)
    };

    \legend{WDC, GDS}

    \end{axis}
\end{tikzpicture}
\caption{Average Precision for WDC and GDS across different feature settings. \textit{'D'} represents distributional features, \textit{'S'} denotes statistical features and \textit{'C'} refers to contextual (headers) features. The results illustrate the performance of these feature combinations for both the WDC and GDS datasets.}
\label{fig:ablation_features_updated}
\end{figure}
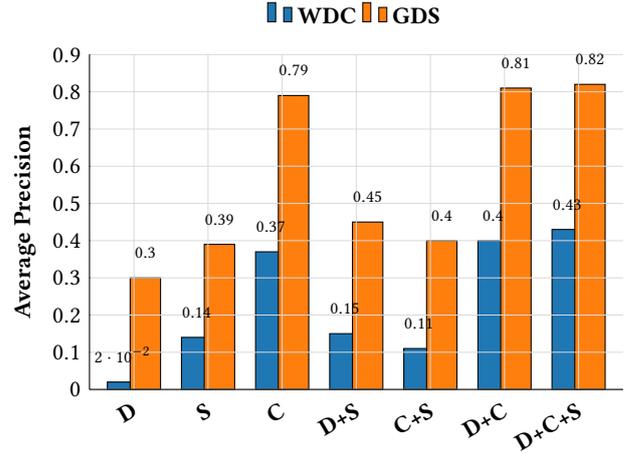

For each combination, we generated an embedding matrix between column pairs and calculated precision by determining how often the top-k most similar columns matched the ground truth labels. This experiment provides insights into the impact of each feature type in accurately detecting column semantics using numerical embeddings. The results, highlighting the performance of each feature combination, are presented in Figure \ref{fig:ablation_features_updated} for both WDC and GDS datasets.

The following can be observed: 
\begin{enumerate}
    \item \textbf{Among the individual feature types in Gem, Contextual performs better than Statistical that performs better than Distributional.} The Contextual features act on column headers (and not values) whereas both Statistical and Distributional features act on column values (and not headers). The results for all three feature types are better in GDS than WDC. The Distributional features, by way of the GMM, are designed to model underlying latent distributions, and work well in cases where the data is naturally segmented into distinct distributions. However, such distributions are not well-defined in GDS and WDC for numerical columns. In such cases, GMM fails to give a full characterisation of a column.

    \item \textbf{Distributional features combine effectively with both Statistical and Contextual features.}  This is reflected in the fact that (D+S) performs better than both D and S independently and that (D+C) performs better than both D and C independently.  This is in contrast with Statistical features, which combine less well with Contextual features; (C+S) performs worse than C on its own for both datasets. Combining Distributional features, by way of the GMM, with Statistical features compensates for their weaknesses in isolation. GMM captures fine-grained distributional details, while statistical features provide broader, high-level insights. Together, they form a more comprehensive representation.

    \item \textbf{All three features together perform better than pairs of features}. Indeed, (D+C+S) performs much better than (C+S) and (D+S), but only slightly better than (D+C).

\end{enumerate}

\subsection{Impact of Gaussian components}
\label{ablation_components}
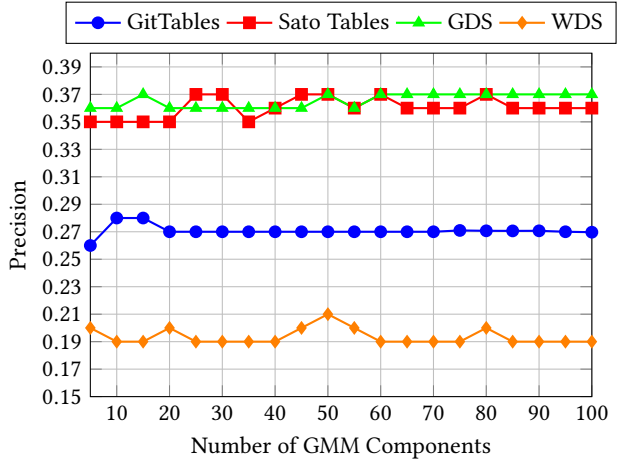
\begin{figure}[t]
    \centering
    \begin{tikzpicture}
        \begin{axis}[
            width=\columnwidth,
            height=0.75\columnwidth,
            xlabel={Number of GMM Components},
            ylabel={Precision},
            legend style={at={(0.5,1.15)}, anchor=north,legend columns=-1},
            grid=major,
            xmin=5, xmax=100,
            ymin=0.15, ymax=0.40,
            xtick={10, 20, 30, 40, 50, 60, 70, 80, 90, 100},
            ytick={0.15, 0.17, 0.19, 0.21, 0.23, 0.25, 0.27, 0.29, 0.31, 0.33, 0.35, 0.37, 0.39},
            scaled ticks=false,
            tick label style={/pgf/number format/fixed},
            legend cell align={left},
            mark options={solid},
        ]
        \addplot[color=blue, mark=*, thick] table[x=n_components, y=GitTables] {
            n_components GitTables SatoTables GDS WDS
            5 0.26 0.35 0.36 0.2
            10 0.28 0.35 0.36 0.19
            15 0.28 0.35 0.37 0.19
            20 0.27 0.35 0.36 0.2
            25 0.27 0.37 0.36 0.19
            30 0.27 0.37 0.36 0.19
            35 0.27 0.35 0.36 0.19
            40 0.27 0.36 0.36 0.19
            45 0.27 0.37 0.36 0.2
            50 0.27 0.37 0.37 0.21
            55 0.27 0.36 0.36 0.2
            60 0.27 0.37 0.37 0.19
            65 0.27 0.36 0.37 0.19
            70 0.27 0.36 0.37 0.19
            75 0.270952477 0.36 0.37 0.19
            80 0.270686352 0.37 0.37 0.2
            85 0.270624407 0.36 0.37 0.19
            90 0.270682688 0.36 0.37 0.20
            95 0.27 0.36 0.37 0.20
            100 0.269678465 0.36 0.37 0.19
        };
        \addlegendentry{GitTables}

        \addplot[color=red, mark=square*, thick] table[x=n_components, y=SatoTables] {
            n_components GitTables SatoTables GDS WDS
            5 0.26 0.35 0.36 0.2
            10 0.28 0.35 0.36 0.19
            15 0.28 0.35 0.37 0.19
            20 0.27 0.35 0.36 0.2
            25 0.27 0.37 0.36 0.19
            30 0.27 0.37 0.36 0.19
            35 0.27 0.35 0.36 0.19
            40 0.27 0.36 0.36 0.19
            45 0.27 0.37 0.36 0.2
            50 0.27 0.37 0.37 0.21
            55 0.27 0.36 0.36 0.2
            60 0.27 0.37 0.37 0.19
            65 0.27 0.36 0.37 0.19
            70 0.27 0.36 0.37 0.19
             75 0.270952477 0.36 0.37 0.19
            80 0.270686352 0.37 0.37 0.2
            85 0.270624407 0.36 0.37 0.19
            90 0.270682688 0.36 0.37 0.20
            95 0.27 0.36 0.37 0.20
            100 0.269678465 0.36 0.37 0.19
        };
        \addlegendentry{Sato Tables}

        \addplot[color=green, mark=triangle*, thick] table[x=n_components, y=GDS] {
            n_components GitTables SatoTables GDS WDS
            5 0.26 0.35 0.36 0.2
            10 0.28 0.35 0.36 0.19
            15 0.28 0.35 0.37 0.19
            20 0.27 0.35 0.36 0.2
            25 0.27 0.37 0.36 0.19
            30 0.27 0.37 0.36 0.19
            35 0.27 0.35 0.36 0.19
            40 0.27 0.36 0.36 0.19
            45 0.27 0.37 0.36 0.2
            50 0.27 0.37 0.37 0.21
            55 0.27 0.36 0.36 0.2
            60 0.27 0.37 0.37 0.19
            65 0.27 0.36 0.37 0.19
            70 0.27 0.36 0.37 0.19
            75 0.270952477 0.36 0.37 0.19
            80 0.270686352 0.37 0.37 0.2
            85 0.270624407 0.36 0.37 0.19
            90 0.270682688 0.36 0.37 0.20
            95 0.27 0.36 0.37 0.20
            100 0.269678465 0.36 0.37 0.19
        };
        \addlegendentry{GDS}
        \addplot[color=orange, mark=diamond*, thick] table[x=n_components, y=WDS] {
            n_components GitTables SatoTables GDS WDS
            5 0.26 0.35 0.36 0.2
            10 0.28 0.35 0.36 0.19
            15 0.28 0.35 0.37 0.19
            20 0.27 0.35 0.36 0.2
            25 0.27 0.37 0.36 0.19
            30 0.27 0.37 0.36 0.19
            35 0.27 0.35 0.36 0.19
            40 0.27 0.36 0.36 0.19
            45 0.27 0.37 0.36 0.2
            50 0.27 0.37 0.37 0.21
            55 0.27 0.36 0.36 0.2
            60 0.27 0.37 0.37 0.19
            65 0.27 0.36 0.37 0.19
            70 0.27 0.36 0.37 0.19
            75 0.270952477 0.36 0.37 0.19
            80 0.270686352 0.37 0.37 0.2
            85 0.270624407 0.36 0.37 0.19
            90 0.270682688 0.36 0.37 0.19
            95 0.27 0.36 0.37 0.19
            100 0.269678465 0.36 0.37 0.19
        };
        \addlegendentry{WDS}

        \end{axis}
    \end{tikzpicture}
    \caption{Performance comparison across different numbers of GMM components for all datasets.}
    \label{fig:ablation_GMM}
\end{figure}

In this section, we assess the impact of the number of GMM components on Gem's performance. We vary the number of GMM components from 5 to 100 across all datasets, and the results are presented in Figure \ref{fig:ablation_GMM}. Our observations indicate that the number of Gaussian components does not significantly impact Gem's overall performance on any of the four datasets. Precision results for GitTables remain consistently around 0.27 to 0.28, with minimal fluctuations as the number of components increases. Similarly, SatoTables show a stable range of 0.35 to 0.37, while GDS consistently remains around 0.36 to 0.37. For WDS, precision scores slightly vary between 0.19 and 0.21, indicating no significant improvement with more Gaussian components. This stability across different numbers of components suggests that Gem's performance is robust to the choice of Gaussian mixture complexity. Gem's precision does not show any significant spikes or drops, reinforcing that increasing the number of Gaussian components beyond a certain point does not contribute to a significant performance gain. This consistency is important as it implies that Gem can achieve reliable results without extensive tuning of the Gaussian component parameter.

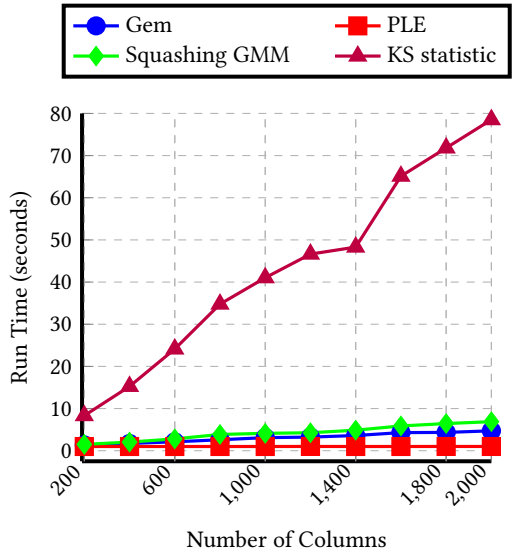
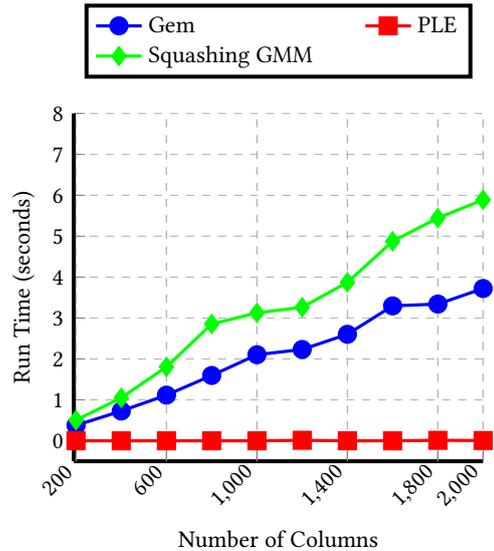
\begin{figure*}[t]
    \centering
    \begin{subfigure}[t]{0.45\textwidth}  
        \centering
        \begin{tikzpicture}
           \begin{axis}[
            width=0.90\columnwidth,
            height=0.80\columnwidth,
            xlabel={Number of Columns},
            ylabel={Run Time (seconds)},
            xmin=190, xmax=2000,               
            ymin=-2.5, ymax=80,                  
            xtick={200,600,1000,1400,1800,2000},
            ytick={0,10,20,30,40,50,60,70,80},
            xticklabel style={rotate=45, anchor=east},
            grid=major,
            grid style={dashed, gray!60},
            legend style={at={(0.5,1.1)}, anchor=south, legend columns=2, /tikz/every even column/.append style={column sep=0.5cm}},
            legend cell align={left},
            axis y line*=left,
            axis x line*=bottom,
            tick pos=left,
            extra x ticks={200,600,1000,1400,1800,2000},
            extra tick style={tick label style={opacity=0}},
            line width=1.5pt
        ]

            \addplot[color=blue, mark=*, mark options={fill=blue}, line width=1.2pt, mark size=3pt] coordinates {
                (200,1.377796888) (400,1.727458715) (600,2.116602182) (800,2.596458197) 
                (1000,3.103818655) (1200,3.232205391) (1400,3.603071213) (1600,4.297484398) 
                (1800,4.341497421) (2000,4.721308947)
            };
            \addlegendentry{Gem}

            \addplot[color=red, mark=square*, mark options={fill=red}, line width=1.2pt, mark size=3pt] coordinates {
                (200,1) (400,1) (600,1) (800,1) 
                (1000,1) (1200,1.012038708) (1400,1) (1600,1) 
                (1800,1.011105061) (2000,1.006778955)
            };
            \addlegendentry{PLE}

            \addplot[color=green, mark=diamond*, mark options={fill=green}, line width=1.2pt, mark size=3pt] coordinates {
                (200,1.504442215) (400,2.050635338) (600,2.806079865) (800,3.8537395) 
                (1000,4.125673056) (1200,4.263442278) (1400,4.870949745) (1600,5.874781132) 
                (1800,6.444587231) (2000,6.8904531)
            };
            \addlegendentry{Squashing GMM}

            \addplot[color=purple, mark=triangle*, mark options={fill=purple}, line width=1.2pt, mark size=3pt] coordinates {
                (200,8.324155569) (400,15.25622654) (600,24.12699962) (800,34.77174211) 
                (1000,41.02783346) (1200,46.63266635) (1400,48.29568577) (1600,65.11139584) 
                (1800,71.81062698) (2000,78.51448226)
            };
            \addlegendentry{KS statistic}

            \end{axis}
        \end{tikzpicture}
        \caption{Run time comparison of Gem with other methods.}
        \label{fig:a}
    \end{subfigure}
    \hspace{1em} 
    \begin{subfigure}[t]{0.45\textwidth}  
        \centering
        \begin{tikzpicture}
           \begin{axis}[
            width=0.90\columnwidth,
            height=0.80\columnwidth,
            xlabel={Number of Columns},
            ylabel={Run Time (seconds)},
            xmin=190, xmax=2000,               
            ymin=-0.5, ymax=8,                 
            xtick={200,600,1000,1400,1800,2000},
            ytick={0,1,2,3,4,5,6,7,8},
            xticklabel style={rotate=45, anchor=east},
            grid=major,
            grid style={dashed, gray!60},
            legend style={at={(0.5,1.1)}, anchor=south, legend columns=2, /tikz/every even column/.append style={column sep=0.6cm}},
            legend cell align={left},
            axis y line*=left,
            axis x line*=bottom,
            tick pos=left,
            extra x ticks={200,600,1000,1400,1800,2000},
            extra tick style={tick label style={opacity=0}},
            line width=1.5pt
        ]

            \addplot[color=blue, mark=*, mark options={fill=blue}, line width=1.2pt, mark size=3pt] coordinates {
                (200,0.377796888) (400,0.727458715) (600,1.116602182) (800,1.596458197) 
                (1000,2.103818655) (1200,2.232205391) (1400,2.603071213) (1600,3.297484398) 
                (1800,3.341497421) (2000,3.721308947)
            };
            \addlegendentry{Gem}

            \addplot[color=red, mark=square*, mark options={fill=red}, line width=1.2pt, mark size=3pt] coordinates {
                (200,0) (400,0) (600,0) (800,0) 
                (1000,0) (1200,0.012038708) (1400,0) (1600,0) 
                (1800,0.011105061) (2000,0.006778955)
            };
            \addlegendentry{PLE}

            \addplot[color=green, mark=diamond*, mark options={fill=green}, line width=1.2pt, mark size=3pt] coordinates {
                (200,0.504442215) (400,1.050635338) (600,1.806079865) (800,2.8537395) 
                (1000,3.125673056) (1200,3.263442278) (1400,3.870949745) (1600,4.874781132) 
                (1800,5.444587231) (2000,5.8904531)
            };
            \addlegendentry{Squashing GMM}

            \end{axis}
        \end{tikzpicture}
        \caption{Zoom view of Gem, PLE, and Squashing GMM to visualize patterns.}
        \label{fig:b}
    \end{subfigure}

    \caption{Run time comparison of different methods: (a) Overall view; (b) Zoomed-in view.}
    \label{fig:runtime_plot}
\end{figure*}

\subsection{Scalability analysis}

 We perform a scalability analysis (see Figure \ref{fig:runtime_plot}) of Gem with PLE, KS statistic, and Squashing GMM to evaluate how each method's runtime scales as the number of columns in the dataset increases from 200 to 3000. For consistency, we measured each method's run time to generate embeddings. For GMM, this involved generating probability-based embeddings, while for PLE and Squashing GMM, we measured the time to compute their respective embeddings. We measured the time to generate the statistical embedding matrix for the KS statistic. To ensure consistency, we measured the runtime for each dataset size five times and calculated the average runtime for each method.

Figure \ref{fig:a} compares the runtime of Gem and baseline methods, whereas Figure \ref{fig:b} provides a zoomed-in view for a more detailed look at the less computationally intensive methods. In Figure \ref{fig:b}, PLE demonstrates a consistently low runtime across all column sizes, maintaining near-zero values and exhibiting a constant trend as the number of columns increases. Gem, in contrast, shows a gradual rise in runtime with the growth in columns, but this increase remains less than linear. A slight fluctuation occurs around 1400 columns, where the runtime briefly slows its rate of increase before continuing its upward trend. Squashing GMM follows a similar upward trajectory to Gem, with a steady, less-than-linear increase in runtime. On the other hand, the KS statistic, as shown in Figure \ref{fig:a}, experiences linear growth in runtime, making it the most computationally expensive method as the number of columns grows.

\begin{table*}[t]
	\centering
	\caption{Clustering results on the GDS and WDC datasets. We compare the performance of two methods, Gem and Squashing\_SOM, on each dataset using ARI and ACC metrics. The best result for each dataset, based on ARI and ACC, is highlighted in bold.}
	\label{tab:4}
	\begin{tabular}{ccccccccccccccccc}  
 \toprule
 & \multicolumn{8}{c}{Gem} & \multicolumn{8}{c}{Squashing\_SOM} \\  
 \midrule
 & \multicolumn{4}{c}{GDS} & \multicolumn{4}{c}{WDC} & \multicolumn{4}{c}{GDS} & \multicolumn{4}{c}{WDC} \\
 \midrule
 & \multicolumn{2}{c}{TableDC} & \multicolumn{2}{c}{SDCN} & \multicolumn{2}{c}{TableDC} & \multicolumn{2}{c}{SDCN} & \multicolumn{2}{c}{TableDC} & \multicolumn{2}{c}{SDCN} & \multicolumn{2}{c}{TableDC} & \multicolumn{2}{c}{SDCN} \\  
 & ARI & ACC & ARI & ACC & ARI & ACC & ARI & ACC & ARI & ACC & ARI & ACC & ARI & ACC & ARI & ACC \\  
Headers only & \textbf{0.69} & \textbf{0.76} & 0.65 & 0.68 & \textbf{0.31} & 0.41 & 0.30 & 0.41 & - & - & - & - & - & - & - & - \\  
Values only & \textbf{0.39} & \textbf{0.48} & \textbf{0.39} & 0.46 & \textbf{0.03} & \textbf{0.12} & \textbf{0.03} & \textbf{0.12} & 0.29 & 0.32 & \textbf{0.31} & \textbf{0.33} & \textbf{0.009 }& \textbf{0.21} & \textbf{0.009} & 0.20 \\  
Headers + Values & \textbf{0.78} & \textbf{0.81} & 0.74 & 0.77 & \textbf{0.33} & \textbf{0.43} & 0.27 & 0.38 & \textbf{0.63} & \textbf{0.70} & 0.58 & 0.61 & \textbf{0.009} & 0.20 & \textbf{0.009} & \textbf{0.21} \\ 
\bottomrule
\end{tabular}
\end{table*}

\subsection{Clustering Results}

We evaluated Gem for an additional downstream clustering task by clustering columns with similar semantics using Deep Clustering (DC) algorithms. We applied SDCN \cite{DBLP:conf/www/Bo0SZL020}, a well-known DC algorithm, and TableDC \cite{DBLP:journals/corr/abs-2405-17723}, which was specifically designed to support clustering in data management tasks.This analysis evaluated how well Gem integrates with the clustering methods. We also compare Gem embeddings with the ones generated through Squashing\_SOM to see how different embeddings affect the clustering performance. In the clustering environment, the distributional embeddings produced by Gem and Squashing\_SOM are an input for the autoencoder in the DC algorithm. The results are shown in Table \ref{tab:4}. We observe the following: 

\begin{enumerate}

\item Gem consistently outperforms Squashing\_SOM for both TableDC and SDCN when considering numerical embeddings in the GDS dataset. For example, TableDC with Gem obtained a higher 0.10 ARI and 0.16 ACC than TableDC with Squashing\_SOM on GDS, while the improvement for SDCN with Gem embeddings is 0.08 on ARI and 0.13 ACC. Squashing\_SOM's preserved topological structures, however, struggled to integrate the rich semantic context from SBERT. On the other hand, Gem, which focuses on modeling numerical distributions using GMM, better integrates the contextual information than Squashing\_SOM.

\item  TableDC outperformed SDCN across both datasets under two experimental configurations: headers-only and headers+values. The noticeable improvement is observed in the GDS dataset, where TableDC achieves a 0.08 increase in ACC using the SBERT with the headers-only setting. This highlights the effectiveness of TableDC in leveraging semantic information when focusing exclusively on column headers. 

\item Gem embeddings alone do not integrate well with TableDC and SDCN. However, contextual integration with column values in TableDC shows better performance than SDCN. For example, TableDC and SDCN perform poorly with a 0.39 ARI using values only on GDS. However, TableDC improves by 0.39 ARI when headers are included compared to SDCN, which improves by 0.35 ARI.

\item Like column embeddings, column clustering has poorer results with the WDC dataset than the GDS dataset for both SDCN and TableDC. This arises from the inherent complexity and overlap in the WDC headers, which leads to ambiguities in both downstream tasks. Additionally, the WDC dataset has more varied and noisy data distributions, making it harder for SDCN and TableDC to cluster similar columns effectively. For example, columns \textit{"journal\_Rank"} and \textit{"Book\_Rank"} have similar ranking values, leading to large clusters in both SDCN and TableDC, which is a mis-classification.

\end{enumerate}

\section{Conclusion}

Numerical data is prominent in tabular datasets, and thus, embeddings for database columns can usefully treat numerical data as a first-class citizen.  To enable this, we propose Gem, which focuses on numerical data through a signature mechanism that generates a probability matrix for each column, indicating the likelihood of belonging to specific Gaussian components.  Experiments have (i) shown that Gem outperforms previous numerical embedding proposals (i.e., \cite{DBLP:conf/nips/GorishniyRB22,DBLP:conf/emnlp/JiangNGCZST20}) for semantic type detection of column using numerical embedding over a variety of datasets; and (ii) shown that Gem embeddings can be combined effectively with other evidence on the semantics of a column, such as column headers, both for column clustering and embeddings.  

\bibliographystyle{ACM-Reference-Format}
\bibliography{sample-base}

\end{document}